\documentclass[useAMS,usenatbib]{mn2e}
\usepackage{graphicx,fleqn}
\DeclareGraphicsExtensions{.eps,.ps,.eps.gz,.ps.gz,.eps.Z}
\title[Timing analysis of RX J0720.4--3125: II]{Timing Analysis of the Isolated
Neutron Star RX~J0720.4--3125 Revisited
\thanks{Based on observations
obtained with \xmm, an ESA science mission with
instruments and contributions directly funded by ESA Member States and the
USA (NASA).}}

\author[]{Mark Cropper$^{1}$, Frank Haberl$^{2}$, Silvia Zane$^{1}$, 
Vyacheslav E. Zavlin$^{2}$ \\
$^{1}$Mullard Space Science Lab, University College London,
Holmbury St. Mary, Dorking, Surrey, RH5 6NT, UK\\
$^{2}$ Max Planck Institut f\"ur Extraterrestrische Physik,
Giessenbachstrasse, D-85748 Garching, Germany}

\date{Received: }

\begin{document}

\newcommand{\dg} {^{\circ}}
\outer\def\gtae {$\buildrel {\lower3pt\hbox{$>$}} \over
{\lower2pt\hbox{$\sim$}} $}
\outer\def\ltae {$\buildrel {\lower3pt\hbox{$<$}} \over
{\lower2pt\hbox{$\sim$}} $}
\newcommand{\ergscm} {ergs s$^{-1}$ cm$^{-2}$}
\newcommand{\ergss} {ergs s$^{-1}$}
\newcommand{\ergsd} {ergs s$^{-1}$ $d^{2}_{100}$}
\newcommand{\pcmsq} {cm$^{-2}$}
\newcommand{\ros} {{\it ROSAT}}
\newcommand{\xmm} {{\it XMM-Newton}}
\newcommand{\exo} {{\it EXOSAT}}
\newcommand{\sax} {{\it BeppoSAX}}
\newcommand{\chandra} {{\it Chandra}}
\def\rchi{{${\chi}_{\nu}^{2}$}}
\newcommand{\Msun} {$M_{\odot}$}
\newcommand{\Mwd} {$M_{wd}$}
\def\Mdot{\hbox{$\dot M$}}
\def\mdot{\hbox{$\dot m$}}
\def\mincir{\raise -2.truept\hbox{\rlap{\hbox{$\sim$}}\raise5.truept
\hbox{$<$}\ }}
\def\magcir{\raise -4.truept\hbox{\rlap{\hbox{$\sim$}}\raise5.truept
\hbox{$>$}\ }}

\maketitle

\begin{abstract} We present a reanalysis of the X-ray data for RX~J0720.4--3125 presented in Zane et al.~(2002), using more data recently
available from \xmm\/ and \chandra. This analysis also corrects the \ros\/ data
used in that paper to the TDB time system, incorporates the revised \xmm\/
barycentric correction available since then, and corrects the definition of the
instantaneous period in the maximum likelihood periodogramme search. However,
we are now unable to find a single coherent period that is consistent with all
\ros, \chandra\/ and \xmm\/ datasets.  From an analysis of the separate
datasets, we have derived limits on the period change of $\dot P = (1.4 \pm
0.6) \times 10^{-13}$~s/s at 99\% confidence level. This is stronger than the
value presented in Zane et al.~(2002), but sufficiently similar that their
scientific conclusions remain unchanged. We examine the implications in more
detail, and find that RX J0720.4--3125 can have been born as a magnetar
provided that it has a young age of $\sim10^4$ yr. A more conservative
interpretation is that the field strength has remained relatively unchanged at
just over $10^{13}$ G, over the $\sim10^6$ yr lifetime of the star.
\end{abstract} 

\section{Introduction}
\label{int}

RXJ0720.4--3125 is a nearby, isolated neutron star (INS) originally 
discovered
with \ros\/ during a systematic survey of the Galactic plane by Haberl et
al. (1997). A clear modulation of the X-ray intensity is detected 
at a period of 8.391 s, which identifies the spin velocity of the neutron 
star. 

RX J0720.4--3125 is, so far, the isolated neutron star with best
studied timing
properties. Its proximity and relatively high brightness made
possible not only to measure the spin period, but also to study
the period changes over a long-term. This is crucial, since the
positive detection of spin-up or spin-down phases can shed light on the
mechanisms that regulate the neutron star interaction with its
surroundings. A large positive spin-down can indicate magneto-dipolar
breaking, leading in turn to an estimate of magnetic field strength and
star spin down age.
In the attempt to measure a secular period change, Zane et al.~(2002, 
hereafter Z02) recently undertook a comprehensive timing analysis of
\ros, \sax, \chandra\/ and \xmm\/ data spanning a total period of $\sim
7$~yrs. Independently, a similar analysis based on an additional
\chandra\/ dataset, but without the \xmm\/ data, was presented by Kaplan
et al. (2002, hereafter K02). Although the two studies agreed in their 
main
conclusion that the period change was less than $\sim 3\times10^{-13}$
s/s, it was evident that there were inconsistencies in the details of the
analysis which have implications for further studies. The arrival times 
computed by the two groups were different. 
K02 also raised
theoretical concerns about the validity of the {\it coherent} analysis of 
the entire dataset carried out by Z02. That was the only method that, at 
that time, permitted the (positive) sign of the period change
to be constrained. In fact, the phase-incoherent analysis presented by 
Kaplan et al.~(2002) still leads to a large uncertainty in $\dot P$, and 
does not permit the discrimination between positive and negative 
spin-down.

The above-mentioned
inconsistencies have motivated us to undertake a complete revision of the
solution. At the time Z02 was published, timing inconsistencies between 
\xmm\/ and radio
data on other pulsars had come to light (Kuster et al.~2001). This was 
traced to
an error in the \xmm\/ Science Analysis System (SAS) task {\it barycen} before
version 13.1 [the spacecraft position vector was constructed as (x, y, y)
instead of (x, y, z)]. Although we argued in Z02 that any effect on the
coherent analysis was small, we nevertheless performed a reanalysis of the
\xmm\/ data to check, and also to investigate whether this corrected the
inconsistencies with the K02 analysis. We found indeed that the effect was
negligible. The implication is therefore that the inconsistencies between the
Z02 and K02 analyses remained.

We have therefore undertaken a complete reanalysis of the dataset to
investigate and eliminate these inconsistencies. This has brought to light that
the times used for the \ros\/ data in both the Z02 and K02 studies are in UTC rather
than TDB as used for \chandra\/ and \xmm; the times of arrival (ToAs) in K02
are incorrectly calculated; and that the formula used to calculate the
instantaneous phase was incorrect in Z02. Substantial new datasets are also available from both \xmm\/ and \chandra\/ which permit a more comprehensive investigation into the period evolution. We report here the results of this
reanalysis.

\section{New Observations and Revisions}

We refer the reader to Z02 and K02 for the details of the previous observations, and
comment here largely only on those aspects which have changed. 
The table of observations is given in Table~\ref{tab:obslog}.

\begin{table*}
\begin{center}
\begin{tabular}{llllrrl}
\hline
Date & Observatory & Instrument & Exposure    & Exposure & Effective & Label \\
     &             &         & identification & duration & exposure  &       \\
     &             &         &                & (s)    & (s)     &       \\
\hline
1990 Oct 11    & \ros      & PSPC & & 178840 &  & RASS \\
1993 Sep 27 & \ros      & PSPC & rp300338n00 &  11980 &  3221 & R93 \\
1994 May 11 & \ros      & HRI & rh201733n00  &  173237 & 7402 & R94 \\
1996 Apr 25 & \ros      & HRI &  rh300508n00 & 7743  &  3125  & R96a \\
1996 May 7  & \ros      & HRI &  rh180100n00 &  7838   & 3566 & R96b \\
1996 Sep 27 & \ros      & HRI &  rh300508a01 &   1498 &  1409 & R96c \\
1996 Nov 3  & \ros      & HRI &  rh400884n00 &  65698 & 33569 & R96d \\
1997 Mar 16 & {\sl BeppoSAX}   & LECS & LECS\_20079001 & 99418 & 17235  & S97    \\
1998 Apr 20 & \ros      & HRI &  rh400944n00 & 460195 &  3566 & R98 \\
2000 Feb 1  & \chandra    & HRC-S(LETG 1st order) & 348+349+745 & 305528 &  37635 & Ch00 \\
2000 May 13 & \xmm & MOS1 + thin filter & 0124100101-001 & 61352 & 61648 & X00a \\
            &       & MOS2 + thin filter & 0124100101-002 & 61648 & 61648 \\
            &       & PN + thin filter   & 0124100101-003 & 52305 & 52305 \\
2000 Nov 21 & \xmm & MOS1 + medium filter & 0132520301-007 & 17997 & 17997 & X00b \\
            &       & MOS2 + medium filter & 0132520301-008 & 17994 & 17994 \\
            &       & PN + medium filter &  0132520301-003 &25651 & 25651 \\
2001 Dec 4  & \chandra    & ACIS-S & 2771--2774  & 171243 & 31532  & Ch01 \\
2002 Nov 6  & \xmm & PN + thin filter   & 0156960201-003, 0156960401-003 & 208582 & 58554 & X02 \\
2003 May 2 & \xmm &  PN + thick filter & 0158360201-023  & 72793 &72793  & X03 \\
\hline
\end{tabular}
\caption{The \ros, \sax, \chandra\/ and \xmm\/
observations of RXJ0720.4--3125 used in this paper.}
\label{tab:obslog}
\end{center}
\end{table*}

\subsection{\xmm\/ Data}

RX J0720.4--3125 has observed four times by \xmm, in revolutions 78, 175,
533/534 and 622. This is a substantial increase over that available to
Z02.

In Z02, for revolution 78 (X00a in Z02), special arrangements
({\it tcs\_fix}) were made to ensure the timing accuracy of the
data in the absence of a full observation data file (ODF). In the intervening period, the observation has been
reprocessed in the \xmm\/ pipeline, so that the standard observation data file
(ODF) is now available. We used SAS 5.2 {\it odfingest} to prepare the ODFs
and were also assisted by U. Lammers at ESA who ran a more recent (not then
public) version. Nothing unexpected with regard to timing could be found in the
logs or in a direct inspection of the ODF time correlation and orbit data. In
running the pipelines (SAS {\it emchain} and {\it epchain}), we tracked down
timing correction extrapolation warnings in EPIC-pn to anomalous auxiliary data
within the first 100 frames after the exposure start in some quadrants, causing
the computation of the time tags to fail. This problem was fixed by U. Lammers
using software which was later released in SAS 5.3.2. Also in EPIC-pn for the
X00b data, we corrected the 1 s errors in the frametime as described in
Z02. For the EPIC-mos data we found some timing warnings referring to non-increasing times: these were flagged automatically and not included in the
later selection procedure we used to create the extracted event lists. In other
regards, the extraction was performed as in Z02. 

The rev. 533/534 observations of RX J0720.4--3125 were made for
calibration purposes on 2002 Nov 6--9. Extending over 3 days, these data
can provide tighter constraints on the period determinations than the
earlier observations. The EPIC-pn data only were extracted using the
procedures described in Z02, with the same extraction parameters. SAS
v5.3.2 was used to process the events. Corrections were applied for the 1
s frametime errors as described in Z02. Rev 622 observations were also
made for calibration purposes on 2003 May 2. These were processed as for rev
533/534.

For both old and new datasets the barycentric correction
was carried out using SAS {\it barycen} version 13.1 and 13.2.

\subsection{\ros\/ Data}

Despite that the \ros\/ data had been processed in two 
different ways in Z02 
(as a double check, 
using both the \ros\/ EXSAS software system and FTOOLs), an error was
made in not converting the time system from UTC to TDB in order to relate
these observations to the \xmm\/ and \chandra\/ data, both of which use
the TT/TDB system. The timings for the \ros\/ data were therefore recomputed, using the {\it convert/utc\_tdb} task in EXSAS. The absolute timing of the 1998 \ros\/ 
data is uncertain, which prevents their use in any coherent 
analysis (Z02).

We have also extracted the data for RX J0720.4--3125 from the \ros\/ All
Sky Survey. These data suffer from relatively low count rates because of
vignetting away from the field centre, and also the obscuration from the
PSPC detector support grids. On the other hand the duration of the
observations is several days (2.07 d) days, so they potentially provide a good
constraint on the period determination. Finally we have also extracted
early HRI observations taken in 1994 May. The source is near the edge of
the detector area, and the observation yields only a small number of
counts. Nevertheless, as for the RASS data, the duration of the
observation provides some useful constraints.

\subsection{\sax\/ Data}

In Z02, the \sax\/ observations were not directly used in the period
analysis, but were folded on the derived periods from the remaining data
to check the consistency of the period solution. As the {\sc ftools} {\it Earth2Sun} correction used in Z02 did not apply the \sax\/ clock correction, the \sax\/ data have been reprocessed using the SAX Data Analysis System and the correction recomputed using the {\it baryconv} task in SAX DAS  (kindly performed by M. Feroci). It was confirmed by T. Oosterbroek (private communication) that the SAX timings after this operation are in the TDB system.

\subsection{\chandra\/ Data}

Since the work of Z02, \chandra\/ observations made in 2001 December 4--6 
have become publically available. These observations were
reported in K02. In this work, we used the Chandra data on RX~J0720.4--3125 
collected in four observations of 2001 December 4--6, with
effective exposures of 15.0, 10.6, 4.1 and 1.9 ks.
The observations were conducted with
the Advanced CCD Imaging Spectrometer (ACIS-S) operated in
Continuous Clocking mode. This observational mode provides
a 2.9 ms time resolution by means of sacrificing spatial
resolution in one dimension. In total, we used 66,822
counts extracted from segments of a 3\farcs44 width in the
one-dimentional images of the source, in the 0.1-1.0 keV range.
The times of arrival were corrected for the dither and the Science
Instrument Module motion as described in detail in Zavlin et al. (2000), 
and transformed to the Solar System
Barycenter using the {\it axBary} tool of the CIAO package\footnote{
{\tt http://asc.harvard.edu/ciao/}}.

\begin{figure*}
\includegraphics[scale=0.4]{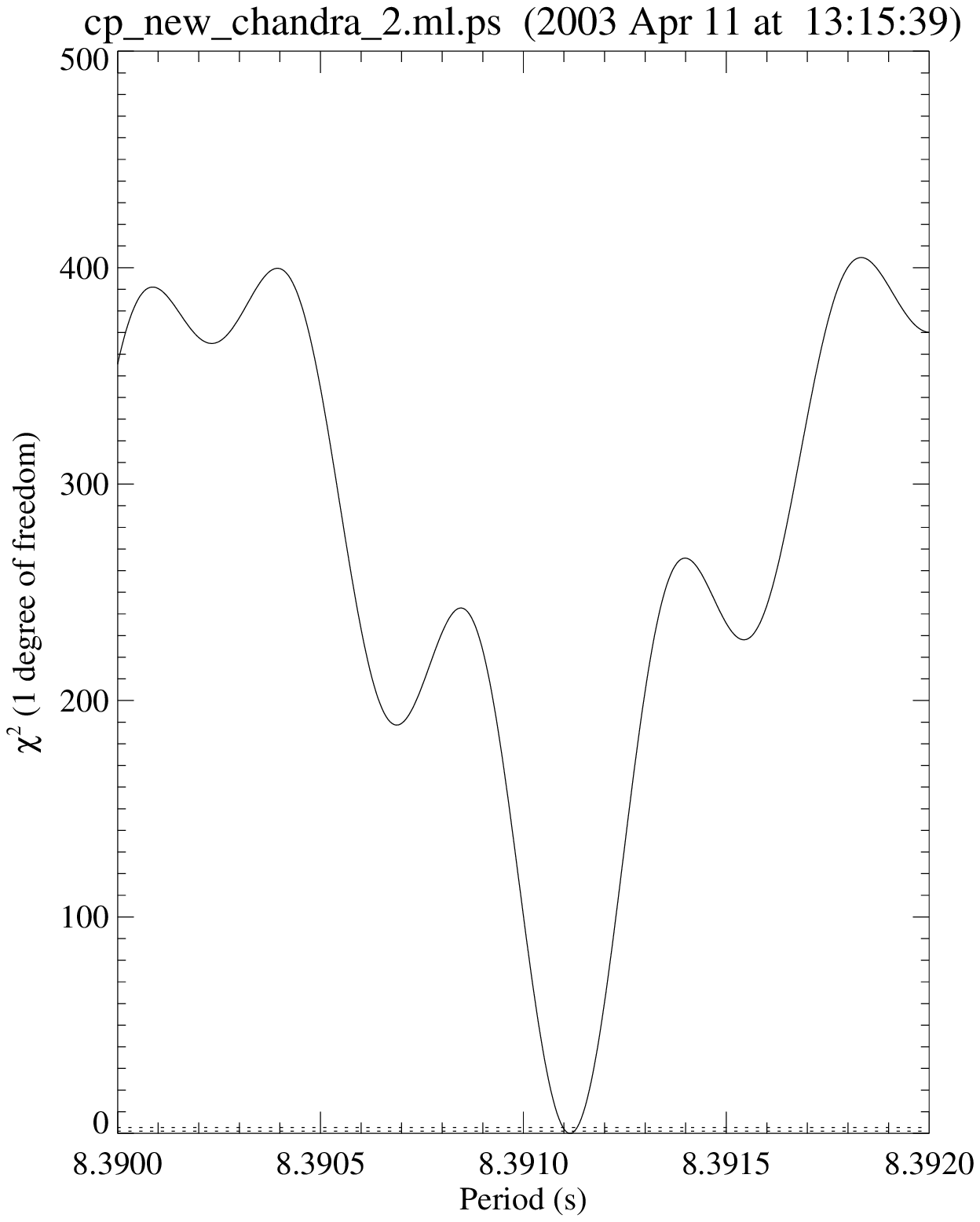}
\hspace*{-10mm}
\includegraphics[scale=0.4]{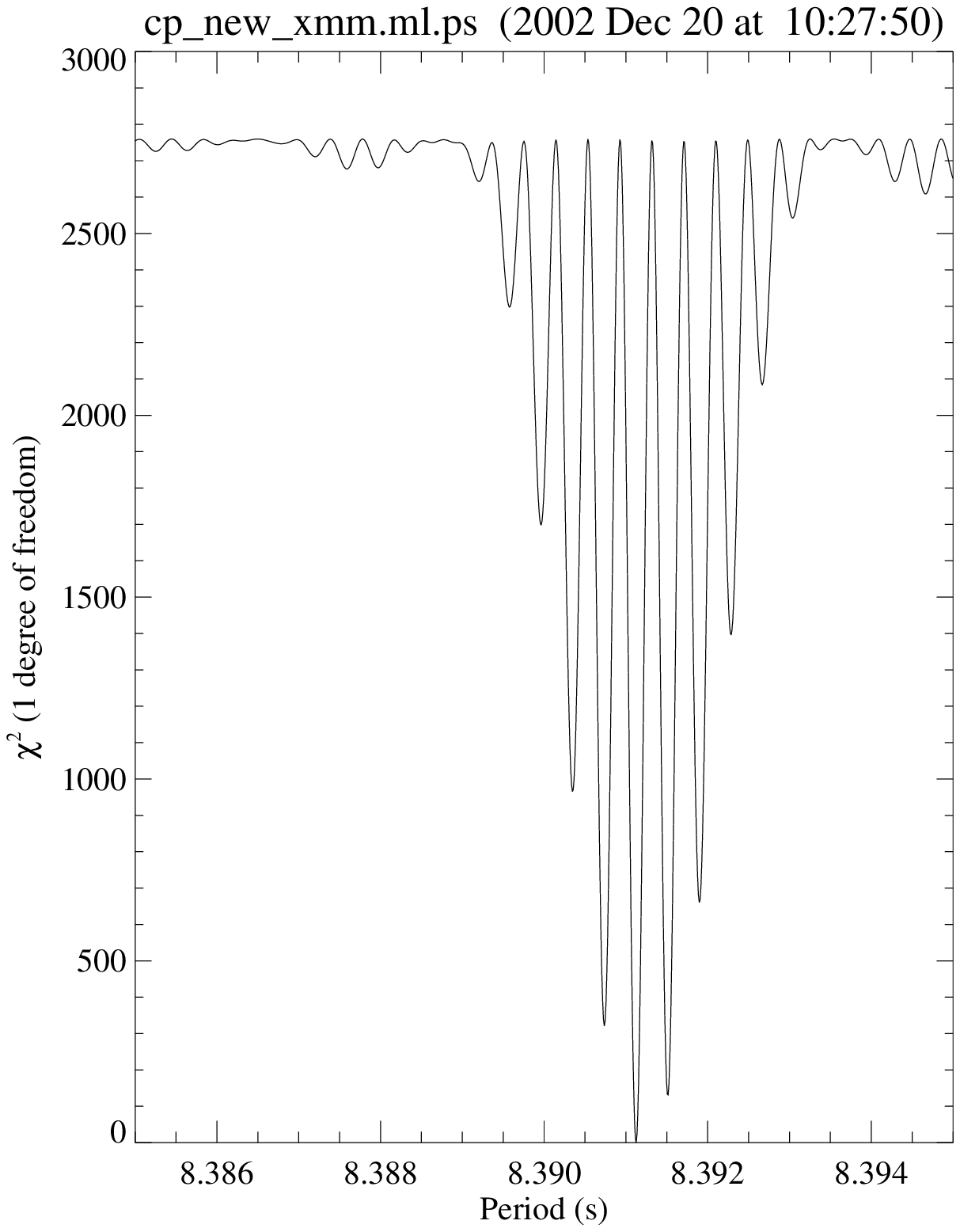}
\hspace*{-10mm}
\includegraphics[scale=0.4]{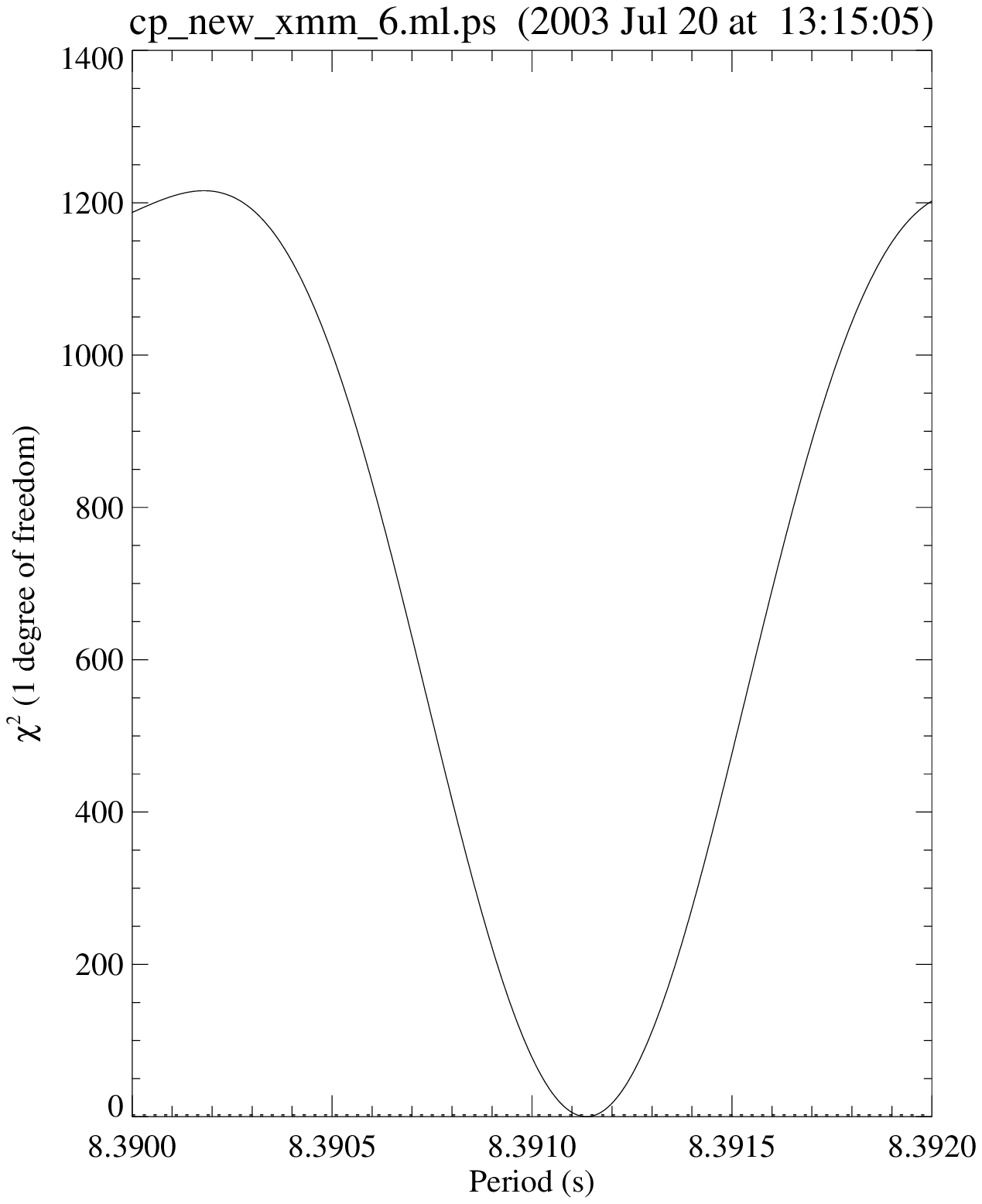}
\caption{The MLPs for the Ch01 data (left), the X02 data (center) and the X03 data (right). The 68 percent $1 \sigma$ confidence levels are set from $\Delta\chi^2=1$, and are not discernible on these plots.}
\label{fig:MLP_individual}
\end{figure*}

\section{Model Revisions}

Z02 (equation A5) defined the instantaneous period to be 
\begin{equation}
P(t) = P_0 + \dot P (t-t_0) \, ,
\label{eq:P_def}
\end{equation}
where $P_0$ is the period at $t = t_0$ and $\dot P$ is the period change. This
period was substituted in the model prediction used in the Maximum Likelihood
Periodogramme (MLP) (equation A4 in Z02)
\begin{equation}
I(t) = a_0[1 + A\cos(2\pi t/P(t) + \theta_0)] \, , 
\label{eq:old_model}
\end{equation}
where $A$ is the normalised amplitude and $\theta_0$ is the phase. 

This substitution is a rough approximation: as the cosine argument in
equation~(\ref{eq:old_model}) is a phase corresponding to the orientation
of the neutron star, the cumulative outcome of the past
history of the period variations is not negligible.

The definition of angular frequency is:  \begin{displaymath} \omega(t) =
\frac{2\pi}{P(t)} = \frac{d\phi}{dt} \, , \end{displaymath} thus,
substituting $P(t)$ from equation~(\ref{eq:P_def}), the phase at time $t$ 
is
\begin{equation} \phi(t)-\phi(t_0) = 2\pi\int^{t}_{t_0}\frac{1}{P_0-\dot P
t_0+\dot P t}dt. \end{equation} In our prescription, $\dot P$ is constant,
so this can be integrated analytically to \begin{equation}
\phi(t)-\phi(t_0) = \frac{2\pi}{\dot P}\ln\frac{P(t)}{P_0} \, .
\end{equation} Accordingly, the model prescription should be revised from
equation~(\ref{eq:old_model}) to 
\begin{equation} I(t) = a_0\left [1 +
A\cos(\frac{2\pi}{\dot P}\ln\frac{P(t)}{P_0} + \theta_0)\right ] \, ,
\label{eq:new_mod}
\end{equation} 
where $P(t)$ is as defined in equation~(\ref{eq:P_def}) and
all of the timing and phase zero points are included in $\theta_0$.

The calculation of the phase in the MLP via equation~(\ref{eq:new_mod}) is 
slower than that previously done via equation~(\ref{eq:old_model}), and 
care is required in order to minimise the
numerical errors resulting from the calculation of the logarithm of a
number close to unity. In particular in the case where $\dot P = 0$, the
solution of equation (3) is $\phi(t) - \phi(t_0) = 2\pi(t-t_0)/P_0$ 
and then the intensity model is simply
\begin{equation}
I(t) = a_0[1+A\cos(\frac{2\pi t}{P_0} + \theta_0)] .
\end{equation}

It could be argued that in the presence of constant torques, a constant
$\dot\nu$ assumption is more valid than a constant $\dot P$. In this
case the accumulated phase is $\phi(t)-\phi(t_0)=2 \pi 
\nu(t)+ 2 \pi \dot\nu 
t^2/2$. In practise we checked that, for $\dot P =5\times10^{-13}$~s/s 
(Z02, 
K02), the accumulated phase difference over ten years between these 
assumptions is
$<0.5$ percent and therefore negligible. The two assumptions give 
consistently similar results, and this also implies that a small second 
derivative term $\ddot \nu $ (of the kind $\ddot \nu =2 \dot P /P^3 - 
\ddot P/P^2$) does not affect the solution. 
With the increase in the time
baseline, however, it will become increasingly possible to distinguish
between the models. In order to facilitate comparison with Z02 and K02 we
retain the constant $\dot P$ model here.

\section{Period Derivative from the Individual Datasets}

The new data allow a more accurate determination of the period derivative
using a conventional least squares fit to the individual periods than was
possible in Z02. This is because the quality of the recent data permit a
breaking of the ambiguities that were present in the dataset used in Z02
caused by the alias patterns in the \ros\/ data. We perform a new
analysis, both in order to provide a robust estimate of $\dot P$ which in
itself is now sufficiently accurate to provide scientific constraints, and
in order to provide a defined region of parameter space for the more
computationally intensive phase-coherent analysis which follows.

\begin{figure}
\begin{center}
\includegraphics[angle=-90,width=\columnwidth]{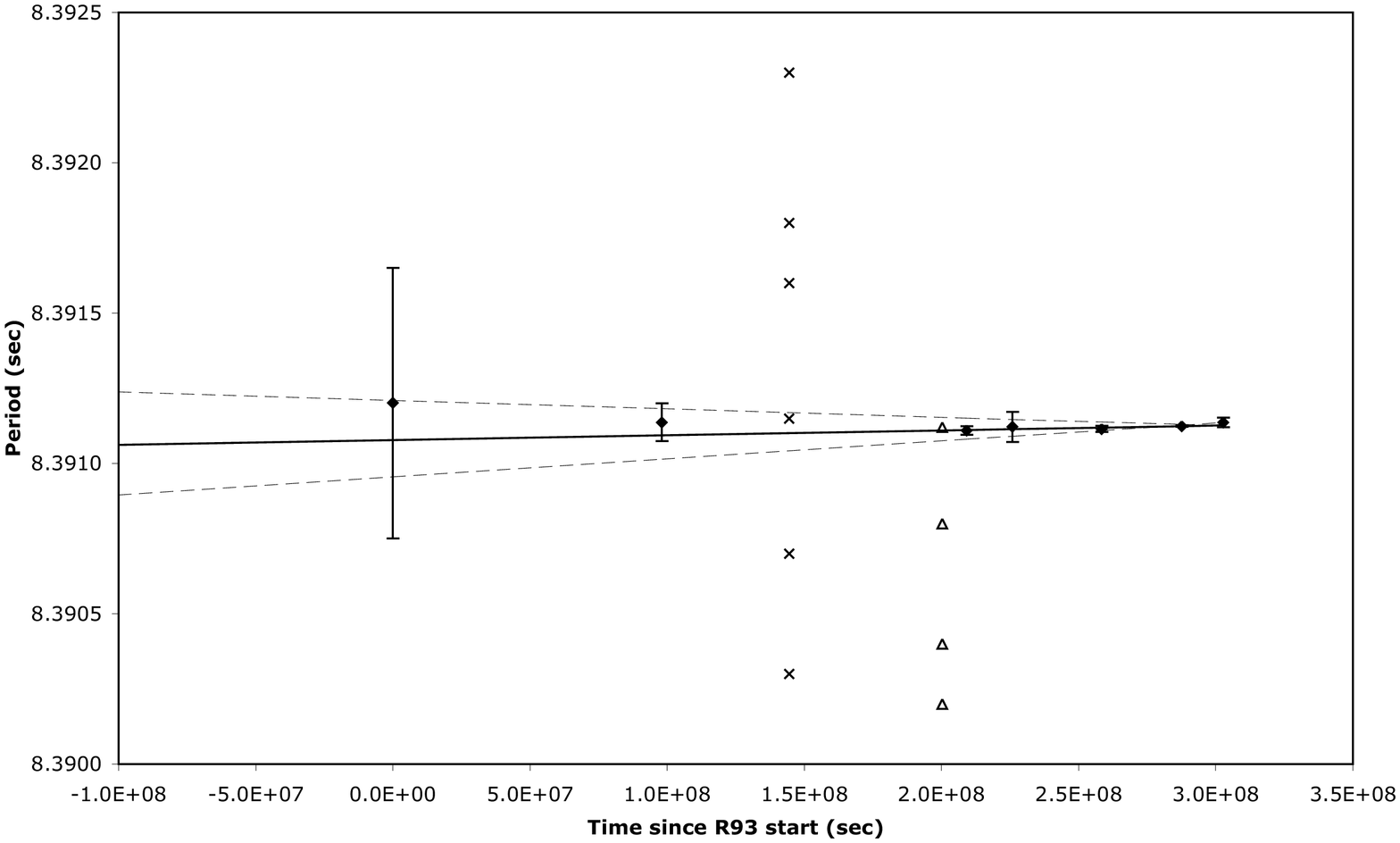}\\
\vspace*{3mm}
\includegraphics[angle=-90,width=\columnwidth]{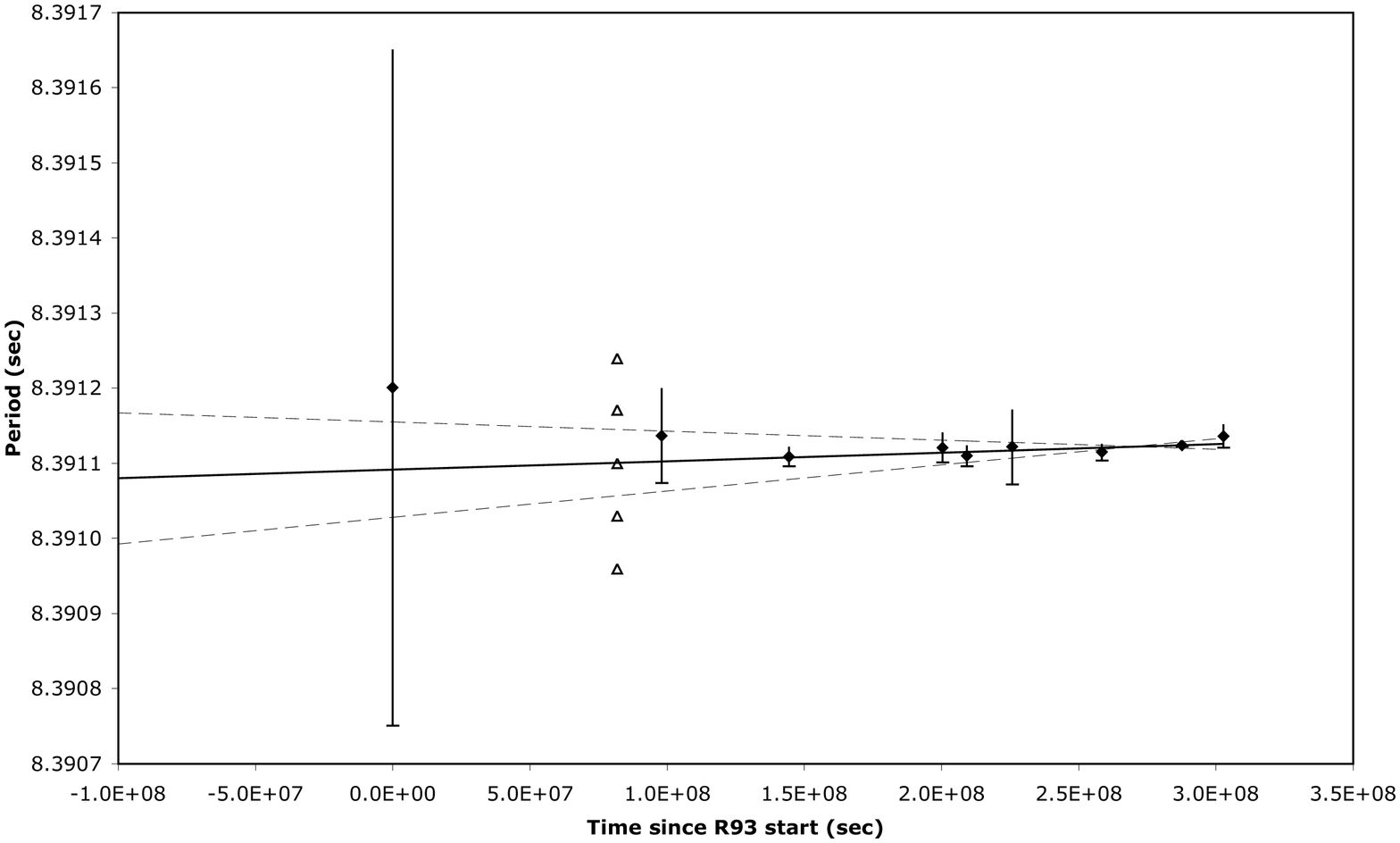}\\
\vspace*{3mm}
\includegraphics[angle=-90,width=\columnwidth]{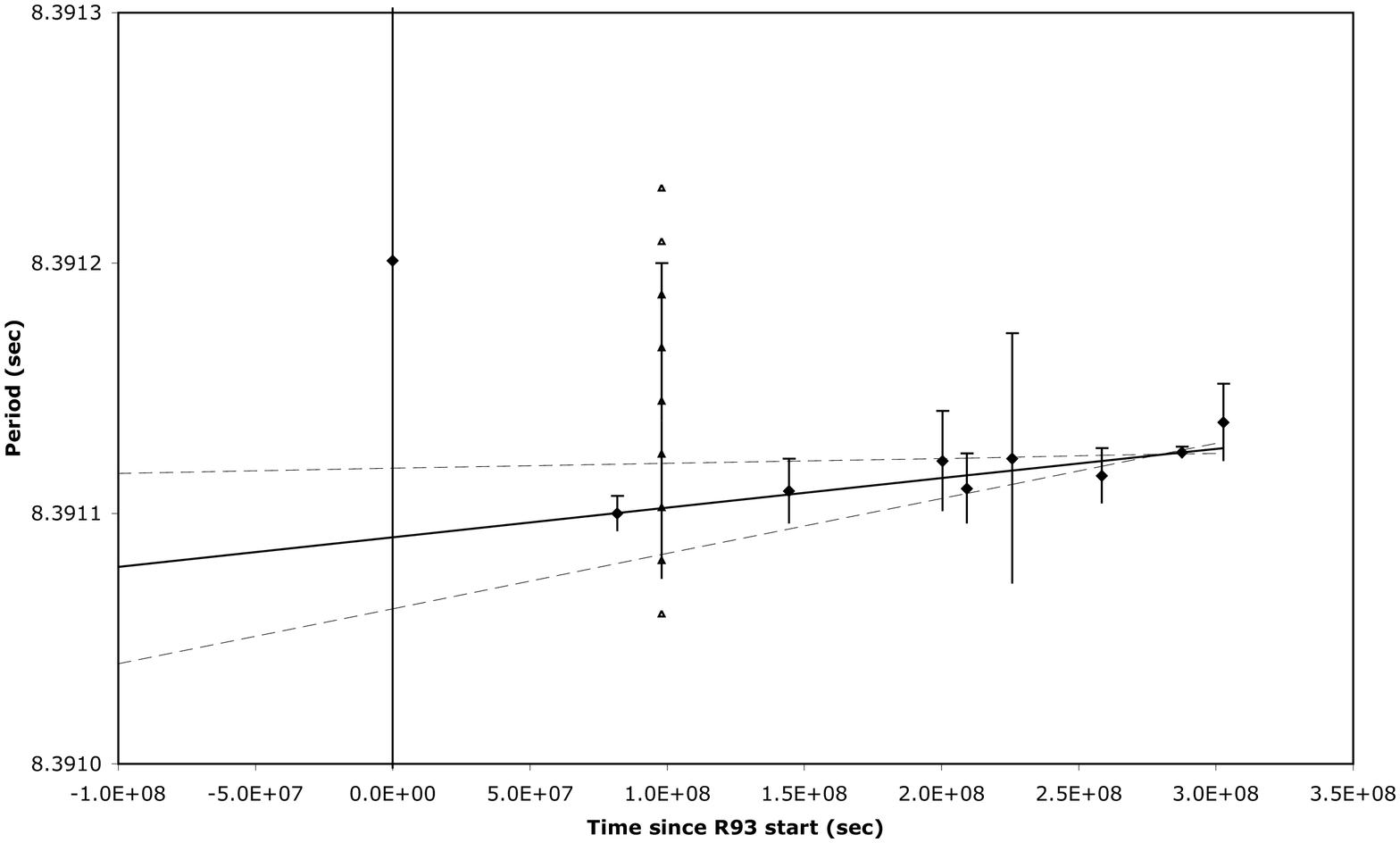}\\
\end{center}
\caption{The periods as derived from the 
MLPs of the \xmm, Ch01, R96d and R93 datasets (Table~\ref{tab:ToA}) shown
as a function of time since the start of the R93 data.  Error bars show
$1\sigma$ uncertainties. The solid line is the best linear fit with dotted
lines showing the 99\% confidence interval. The period aliases for the R98
and Ch00 data are also shown. Only a single alias peak is admissable in
each case (top). The best linear fit including the selected R98 and Ch00
aliases. Also shown are the period aliases from the combined R96a and R96b
data. Only a single alias peak is admissable (centre). The best fit
including the selected aliases from the R96a and R96b data. Also shown are
the period aliases from the combined R96c and R96d data. Again, only a
single alias peak is admissable within the 99\% confidence range (bottom).
}
\label{fig:linfits_P}
\end{figure}

Given the ad-hoc nature of the epochs at which observations have been taken, it
is necessary to proceed in a step-wise fashion. First the best fitting periods
and their uncertainties have been derived for the new \chandra\/ (Ch01) and
\xmm\/ (X02, X03) datasets (Figure~\ref{fig:MLP_individual}), and also the S97
and the RASS datasets using a MLP as in Z02. These were added to those derived
in Z02. The full list is given in Table~\ref{tab:ToA}. The first stage is to
use the \xmm, Ch01, R93 and R96d datasets to determine unequivocally the
appropriate alias peaks in the Ch00 data and the R98 data,  since in these
individual datasets there is no question of the period determination. We
performed a weighted least squares fit using these data, with parameters $P_0$
and $\dot P$ and the extrema of the 99 percent confidence ellipsoids
determined. The result is shown in Figure~\ref{fig:linfits_P} (top). The solid
line is the best fit $P=P_0+\dot P t$, while the dotted lines bound the range
in $P$ permitted from the 99 percent extrema. Also shown are the aliases of the
period determinations for the R96 and Ch00 data. It is clear that only the
aliases corresponding to $P\sim8.3911$ s are consistent with the least squares
fit. These indeed correspond to those selected in Z02 and K02, but this (at
least in Z02) relied on the correspondence of these peaks in the R98 and Ch00
MLPs. 

\begin{figure}
\begin{center}
\includegraphics[scale=0.40,angle=90]{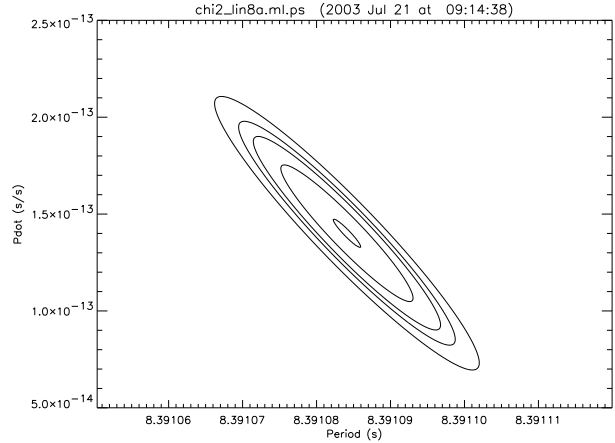}\\
\end{center}
\caption{The 99\%, 95\%, 90\% and 68\% confidence interval contours in the $[P_0, \dot P]$ plane for the data in Figure~\ref{fig:linfits_P} (bottom), and including the selected alias from the R96c and R96d combined data. The central contour is drawn only to identify the minimum in the $\chi^2$ plane. }
\label{fig:linfits_cont}
\end{figure}

It is now possible to include the periods for these peaks in the least squares
fit and repeat the process. The result is shown in Figure~\ref{fig:linfits_P}
(middle). Also shown are the aliases of the combined R96a and R96b
determinations: these are short runs but have a convenient two-week
separation. Again from the 99 percent confidence range, the central alias at
$P\sim8.3911$ s is the only acceptable period. The 99 percent confidence
interval changes slightly if we include the RASS period determination, but the
selection of the aliases is unaffected. This in turn can be used to further
constrain the fit, and to determine which alias should be selected from the
combined R96c and R96d data, which have a 1 month separation. This is shown in
Figure~\ref{fig:linfits_P} (bottom), where the 99 percent confidence limits
identify the alias at 8.391095 s as the appropriate period. Again, the RASS
data can be included without any change to the alias selection. Including the
period determinated from this alias, the least confidence intervals in the
$P_0, \dot P$ plane are shown in Figure~\ref{fig:linfits_cont}.

\begin{table}
\begin{center}
\begin{tabular}{ll@{\hspace*{0mm}}ll@{\hspace*{0mm}}l}
\hline
Dataset &  Period & & ToA  &     \\
               & (s)   & & (MJD) &         \\
\hline
RASS &  8.390935 & $\pm0.000055$ & 48176.0202551& $\pm0.0000059$ \\
R93 &  8.39120 & $\pm 0.00045$ & 49257.2547035 & $\pm0.0000028$ \\
R94 &  8.39065 & $\pm 0.00005$ &49484.0535939 & $\pm0.0000092$ \\
R96a  &  8.3913 & $\pm0.0008 $ & 50198.6873653 & $\pm0.0000031$  \\
R96b  & 8.3925 & $\pm0.0015$ & 50210.5563095 & $\pm0.0000058$ \\
R96a+b  & 8.391100 & $\pm0.000007$ & 50204.6220263 & $\pm0.0000039$  \\
R96c &  8.392 & $\pm0.025$ & 50353.9975527 & $\pm0.0000064$ \\
R96d & 8.391137 & $\pm0.000063$ & 50391.3007509 & $\pm0.0000016$ \\
R96c+d &  8.391095 & $\pm0.000005$ & 50372.8348286 & $\pm0.0000017$ \\
S97 &  8.39113 & $\pm0.00012$ &  50523.7056342 & $\pm0.0000040$ \\
R98 &  8.391109 & $\pm0.000013$ & 50925.6882151 & $\pm0.0000036$ \\
Ch00 &  8.391121 & $\pm0.000020$ & 51577.0395669 & $\pm0.0000026$ \\
X00a {\sc pn} &  8.391110 & $\pm0.000014$ & 51677.4067206 & $\pm0.0000003$ \\
X00a {\sc mos1}  & 8.391110 & $\pm0.000047$ & 51677.4688770  & $\pm0.0000010$ \\
X00a {\sc mos2}  &  8.391113 & $\pm0.000034$ & 51677.4712089 & $\pm0.0000007$ \\
X00b {\sc pn}  & 8.391122 & $\pm0.000050$ &51869.9571032 & $\pm0.0000006$ \\
X00b {\sc mos1}  &  8.391060 & $\pm0.000170$ & 51869.9949797 & $\pm0.0000012$ \\
X00b {\sc mos2}   & 8.390850 & $\pm0.000190$ & 51869.9949796 & $\pm0.0000014$ \\
Ch01  & 8.391115 & $\pm0.000011$& 52248.6768196 & $\pm0.0000008$ \\
X02 {\sc pn} &  8.391124 & $\pm0.000002$ &52585.9688278 & $\pm0.0000003$ \\
X03 {\sc pn} &  8.391136 & $\pm0.000015$ & 52761.9950590 & $\pm0.0000004$ \\
\hline
\end{tabular}
\caption{The periods and ToAs derived from the individual datasets.}
\label{tab:ToA}
\end{center}
\end{table}

The outcome of this analysis is that the 99 percent confidence interval
for the $\dot P$ term lies in the range $0.70\times10^{-13}$ to
$2.10 \times 10^{-13}$ s s$^{-1}$, with the 68 percent interval being
$1.05\times10^{-13}$ to $1.75\times10^{-13}$ s s$^{-1}$. This is
consistent with (but more accurate than) the incoherent analyses by 
Z02 and K02, and establishes that RX J0720.4--3125 is {\it spinning down}.

\section{Period Derivative from a Coherent Analysis}

The change to the model used in the MLP noted above, together with the revised \ros\/
timings (unimportant in the incoherent analysis above) means that the coherent period analysis in section 4.3 of Z02 must be revised -- and indeed the 99 percent confidence limits in Figure~\ref{fig:linfits_cont} exclude all  of the four identified $P_0,\dot P$ pairs in their table 3. 

\subsection{Individual Satellites}
\label{sec:indiv_sats}

As explained in Z02, the time taken to compute coherent 2-dimensional MLPs for a dataset of this size is prohibitive, so that it has to be performed in stages. This can be done in several ways. Here  we have analysed the \ros, \chandra\/ and \xmm\/ datasets separately to start with. This avoids any residual difficulties in relating the different time systems used for the different satellites, or from any different assumptions in the data reduction software. 

In each case we have searched the $[P_0, \dot P]$ plane in the region within the
99\% confidence intervals in Figure~\ref{fig:linfits_cont}, using the
2-dimensional MLP as described in Z02 and modified as described in Section~3
above. Care was taken to ensure that the region was searched on a grid finer
than the Nyquist frequencies to ensure complete sampling of the $\chi^2$ plane.
In the case of the \ros\/ data, we have included the 1993, 1994 and 1996 data,
with a total timespan of 3 years, similar to that for the \xmm\/ data, while the
two \chandra\/ observations span $\sim20$ months. This resulted in MLPs with
dimensions of $2000\times2000$, with datasets of $1.4\times10^6$, $8\times10^4$
and $2\times10^4$ events in the case of \xmm, \chandra\/ and \ros\/
respectively.

\begin{figure}
\begin{center}
\includegraphics[trim=0.1cm 0 0 0,angle=90,width=\columnwidth]{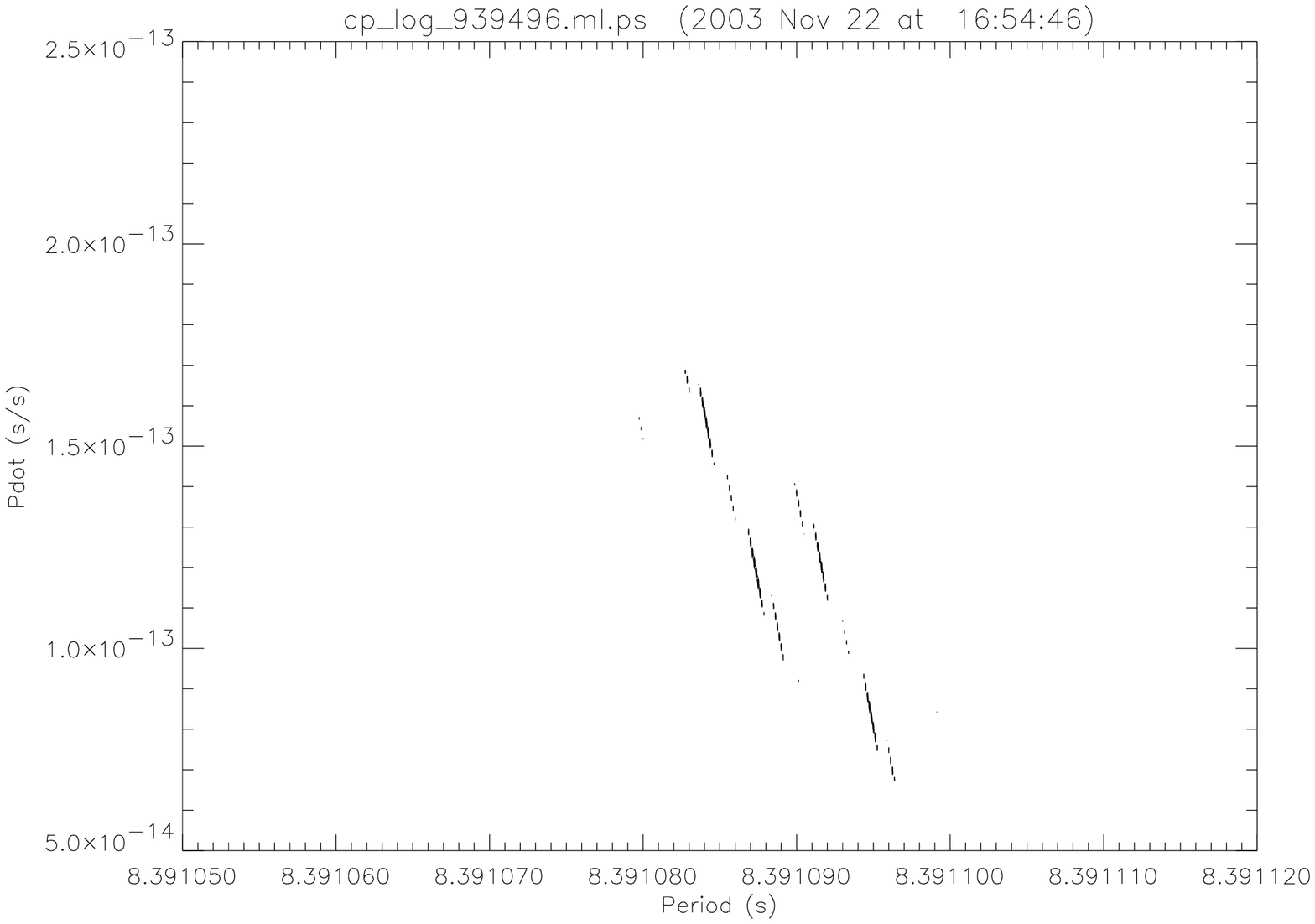}\\
\includegraphics[trim=0.1cm 0 0 0,angle=90,width=\columnwidth]{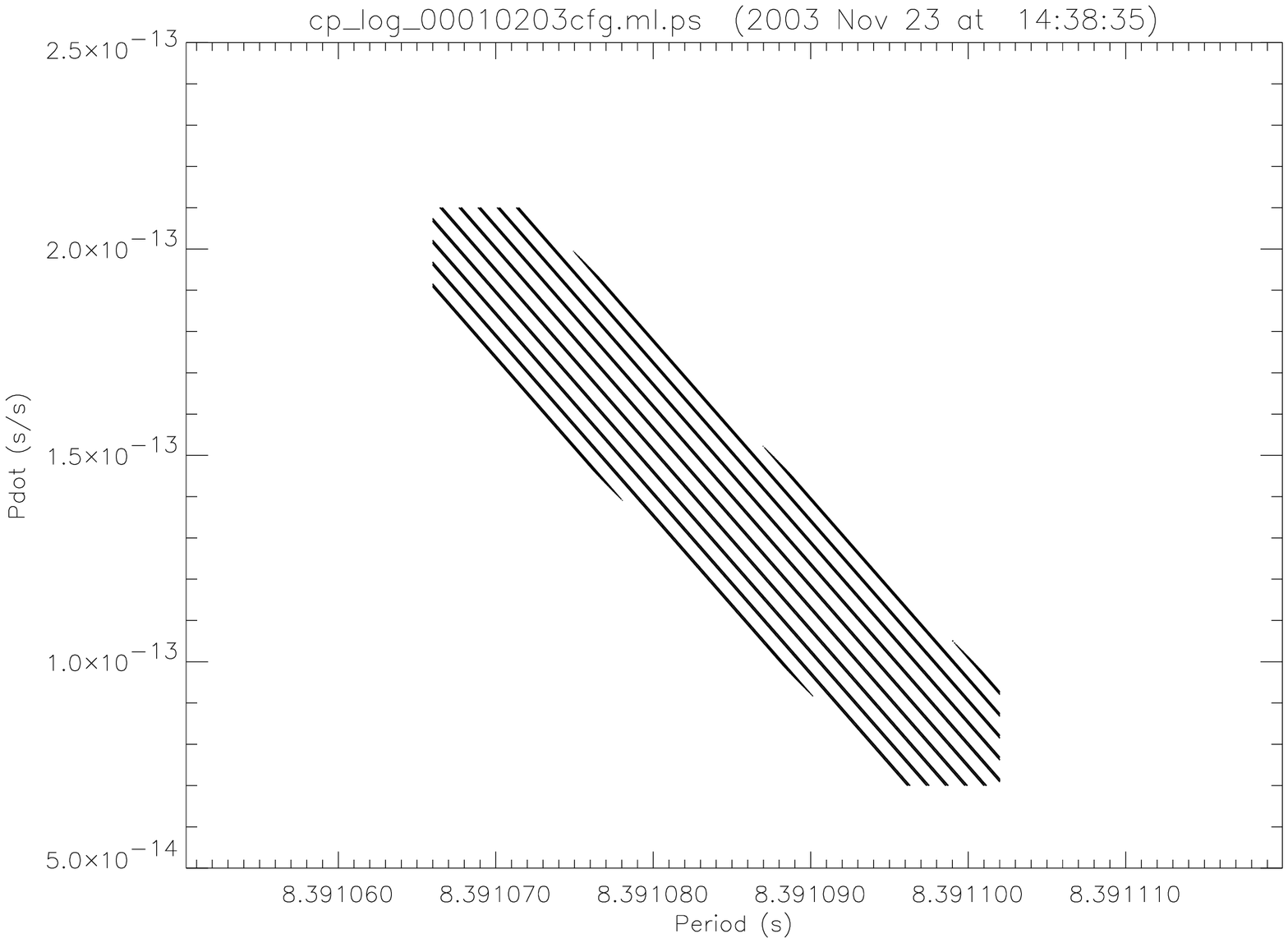}\\
\includegraphics[trim=0.1cm 0 0 0,angle=90,width=\columnwidth]{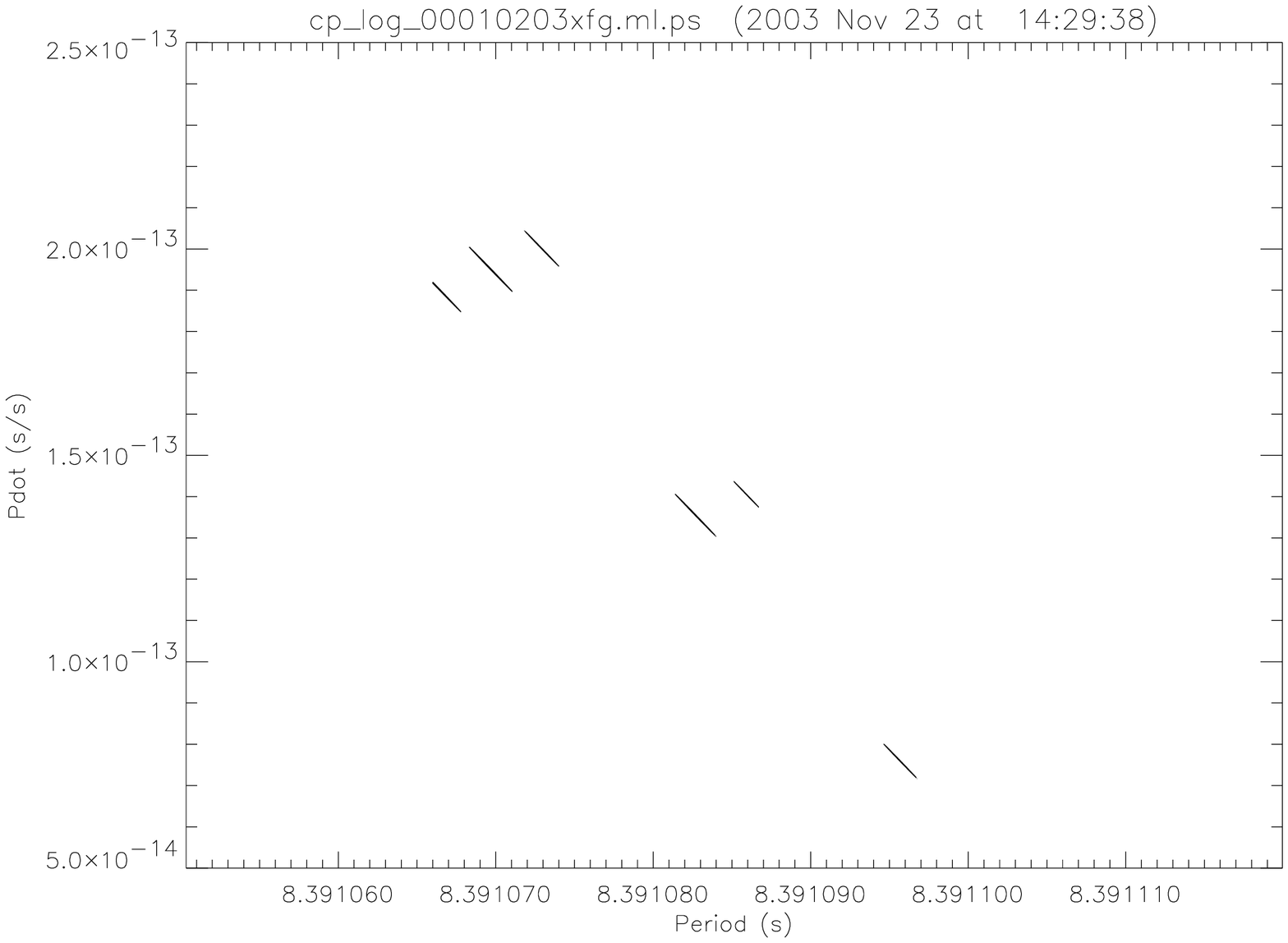}\\
\end{center}
\caption{Contours of $\Delta\chi^2$ in the $[P_0,  \dot P]$ plane for the R93, R94 and R96 dataset (top) for the Ch00 and Ch01 dataset (centre) and the X00, X02 and X03 dataset (bottom). In the \xmm\/ data, were the 99\% confidence interval contours (rather than the 99.9\% in this case) only to be shown, then only the central region at the top left is significant.}
\label{fig:coherent_separated}
\end{figure}

The results are shown in Figure~\ref{fig:coherent_separated}. The outer contours
are the 99\% confidence intervals for two parameters of interest, except for the
\xmm\/ MLP where contours to include 99.9\% significance are shown for
illustration. The \chandra\/ MLP shows diagonal alias patterns resulting from
the availability of data at only two epochs. The \ros\/ MLP consists of two main
diagonals, broken into regions of better and less good fit, together with a
lower likelihood contour parallel to these at shorter periods. In the case of
the \xmm\/ MLP there are five regions, although at the 99\% confidence level,
only the central upper left region is significant.

Overlaying these MLPs, it is clear that there is no single location at which all three datasets intersect. The consequence is that there is an inconsistency between the three determinations. In particular, the $[P_0, \dot P]$ contours from the \xmm\/ MLP do not intersect with those of either the \ros\/ or \chandra\/ MLPs. This has persisted, despite an exhaustive reassessment of the analysis.
 
\subsection{Simulated data}

In order to investigate the viability of the coherent analysis (K02
questioned the applicability of approach in the study by Z02) and to 
verify more stringently the
analysis process, we have generated simulated data with the same mean
levels, noise characteristics and observation durations as the real data
in Table 1. For those observations which were broken into several shorter
sections (such as the \chandra\/ and the \ros\/ data), the simulated data
were generated similarly. An intensity variation of linearly increasing
period, with $P_0$ and $\dot P$ similar to that deduced from the
incoherent analysis was introduced into the data. The simulated dataset
therefore closely matched the real dataset, with the difference that the
$P_0$ and $\dot P$ were known, and the time reference for all the data was
known with certainty to be common. These data were phase folded using
equation~(4) and the known $P_0,\dot P$, and it was checked that all data
co-phased correctly.

We proceeded with the analysis as follows. The $[P_0,\dot P]$ plane
bounded by $[8.391050-8.391120, 0-2\times10^{-13}]$ was searched using the
MLP, first coherently within the Ch00 and X00a,b datasets, then in the
Ch01 and X02a,b datasets and finally in the R96 datasets. Each search
resulted in a diagonal set of contours in $\Delta\chi^2$, corresponding to
the alias patterns. We verified that the known $[P_0,\dot P]$ lay on one
of the best-fit valleys. The orientation of the contours is increasingly
angled with time, so that overlaying the three 2-dimensional MLPs
identified a limited number of locations where the three datasets gave
best fits. We confirmed this more quantitatively by a simple
multiplication of the three MLPs. One of these loci (but not quite the
minimum) was the input $[P_0,\dot P]$. We then carried out a coherent MLP
analysis of all of the R96, Ch00, X00a,b, Ch01, and X02a,b datasets in the
limited $[P_0,\dot P]$ spaces around these loci. This coherent analysis
generated the $\Delta\chi^2$ values requiring mutual co-phasing of all of
the datasets, and now the minimum in the plane was found indeed to
coincide with the input $[P_0,\dot P]$.

This exercise verified both the MLP analysis routines in as realistic fashion as possible, and the  feasibility of identifying the correct period amongst the alias patterns, given the nature of the particular datasets at hand. 

We repeated this order of analysis for the real datasets. Overlaying the MLPs identifies three loci where there is overlap in the patterns, around $[8.3910771, 1.796\times10^{-13}]$, $[8.3910954, 9.50\times10^{-14}]$ and $[8.3910865,  1.3825\times10^{-13}]$. A detailed search using the entire dataset identifies the region around the first solution as the only region where a coherent phase can be maintained over the entire dataset. The best fitting $[P_0, \dot P] = [8.39107712, 1.7956\times10^{-13}]$. However, when folded, this $[P_0, \dot P]$ pair is inconsistent with the X03 dataset.

\subsection{Possible origins for the discrepancy}

At this stage we are not able to arrive at a consistent coherent solution
for a $[P_0, \dot P]$ pair from our extensive dataset. However: 
firstly, we have
shown from the simulations above that the dataset {\it in principle}
contains sufficient information to identify the correct solution, even if
we do not include data from the X03 observations, or from S97, R93 or RASS
datasets. Therefore, if a solution with $\dot P$ constant does exist, the 
technique based on the coherent analysis allows us to identify it. 
Secondly, we have also verified that the software used to
calculate the MLP returns the input parameters for the simulation
correctly: we believe that any coding errors are therefore 
unlikely. Thirdly, in our analysis in Section~\ref{sec:indiv_sats} we have 
taken
care not to mix the data from different satellites, so we have avoided any
residual difficulties in relating the different time systems used for the
different satellites.  

Consequently we conclude that either the event timings used in our data
are incorrectly assigned with respect to observations at other epochs with
the {\it same} satellite, or that the model we have used for a constant
period change [equation (5)] is inappropriate. We discuss these in turn.

We have cross-checked the event timings for our data where they overlap with
K02. K02 have revised their table 1 post-publication, in the version held on the
Astro-ph server which corrects the time system reference and other errors used
to calculate their ToAs in their published version. We have carried out a
detailed cross-check between the ToAs in their revised table and our datasets in
Table 1. Our ToAs are given in Table~\ref{tab:ToA}. For those datasets in common
(R93, R96d, S97, Ch00 and Ch01), the ToAs we calculate match those derived by
K02 closely (to within 2\% of the period), and we derive similar uncertainties. This indicates that these
independently-derived datasets have been reduced consistently, and the
likelihood of systematic errors remaining is now small. However, it is
impossible to eliminate the possibility of error entirely since similar
misconceptions could be resident in the software, or use of the software, for
reducing the data -- such as was the cause of the error in relating \ros\/ and
\chandra\/ data in Z02 and K02.

In the case of the \xmm\/ timings, no such cross-check is currently
possible. It is also of some concern that the best fit $[P_0, \dot P]$
pair from the \xmm\/ observations alone (top centre contours in
Figure~\ref{fig:coherent_separated}) are located outside of the 90\%
confidence intervals calculated from the incoherent analysis
(Figure~\ref{fig:linfits_cont}). We have therefore investigated the timing
accuracy of \xmm\/ in further detail. After the \xmm\/ SAS {\it barycen}
task was corrected for version 13.1, an absolute time shift in Crab pulsar
data of $\sim1.2$ msec remained with respect to the radio data. This
resulted from an incorrect sign used to compensate for internal electronic
delays. Becker (private communication) reports that there is still a time
difference of between 600 and 300 $\mu$sec in the Crab data (apparently
decreasing with time) so minor timing discrepancies remain.

These discrepancies appear to be on such a scale that they would not
affect our \xmm\/ analysis. However, there is some indication that there
may be larger scale discrepancies inherent in some \xmm\/ datasets. These
have arisen in some eclipsing binary data, for example OY Car (Wheatley \&
West 2003), and in EP Dra (Bridge 2004). In the former case, the eclipse
occurs 55 sec early, while in the latter it occurs 69 sec late. These
residuals are significantly larger than permitted by the uncertainties in
the ephemerides for the two stars and would invalidate the coherent
analysis for our \xmm\/ datasets of RX J0720.4--3125. Schwope et al. (2004) also
find some timing discrepancies for DP Leo, but report on the other hand
that the eclipses in \xmm\/
observations of HU Aqr are consistent with the orbital ephemeris. The
situation remains unclear. 

Regarding the appropriateness of a constant $\dot P$ model [equation (5)],
we note three possibilities: glitching may have occurred, or there may be
a deviation from a constant spin down, or a modulation in arrival times
due to the star being in a binary system. Glitching behaviour 
has been observed in neutron stars with spin periods as long as that of
RX~\,J0720.4--3125, as for example in the anomalous X-ray pulsars 
(AXP) 1E2259.1+586
and 1RXS~1708--4009. Long term monitoring of several AXPs with the {\it 
Rossi X-ray Timing Explorer} (Kaspi et al.~2001, Gotthelf et al.~2002, 
Gavriil \& Kaspi~2002) revealed the diversity in the behavior of the 
single objects, ranging from high stability (in 1E2259.1+586 in 
quiescence, and 4U 0142+61 for which a linear fit with constant $\dot \nu$ 
phase-connects data collected over more than 4 years) to instabilities so 
severe that phase-coherent timing is not possible (as in 1E 1048.1--5937). 
In the case of AXPs, timing stability decreases with increasing 
$\dot \nu$ and the frequency derivative of RX J0720.4--3125 is one order of 
magnitude lower than that of the most stable AXPs. However, since the two 
sources belong to different classes of pulsars, it is not really obvious 
how much the extrapolation of this trend can be trusted. 

So far, regrettably the sampling of the current dataset is not likely to 
be sufficient to distinguish whether either of these possibilities is the
explanation for the inconsistencies we obtain in our MLPs.

\subsection{Combined analysis}

\begin{figure}
\begin{center}
\includegraphics[trim=0.1cm 0 0 0,angle=90,width=\columnwidth]{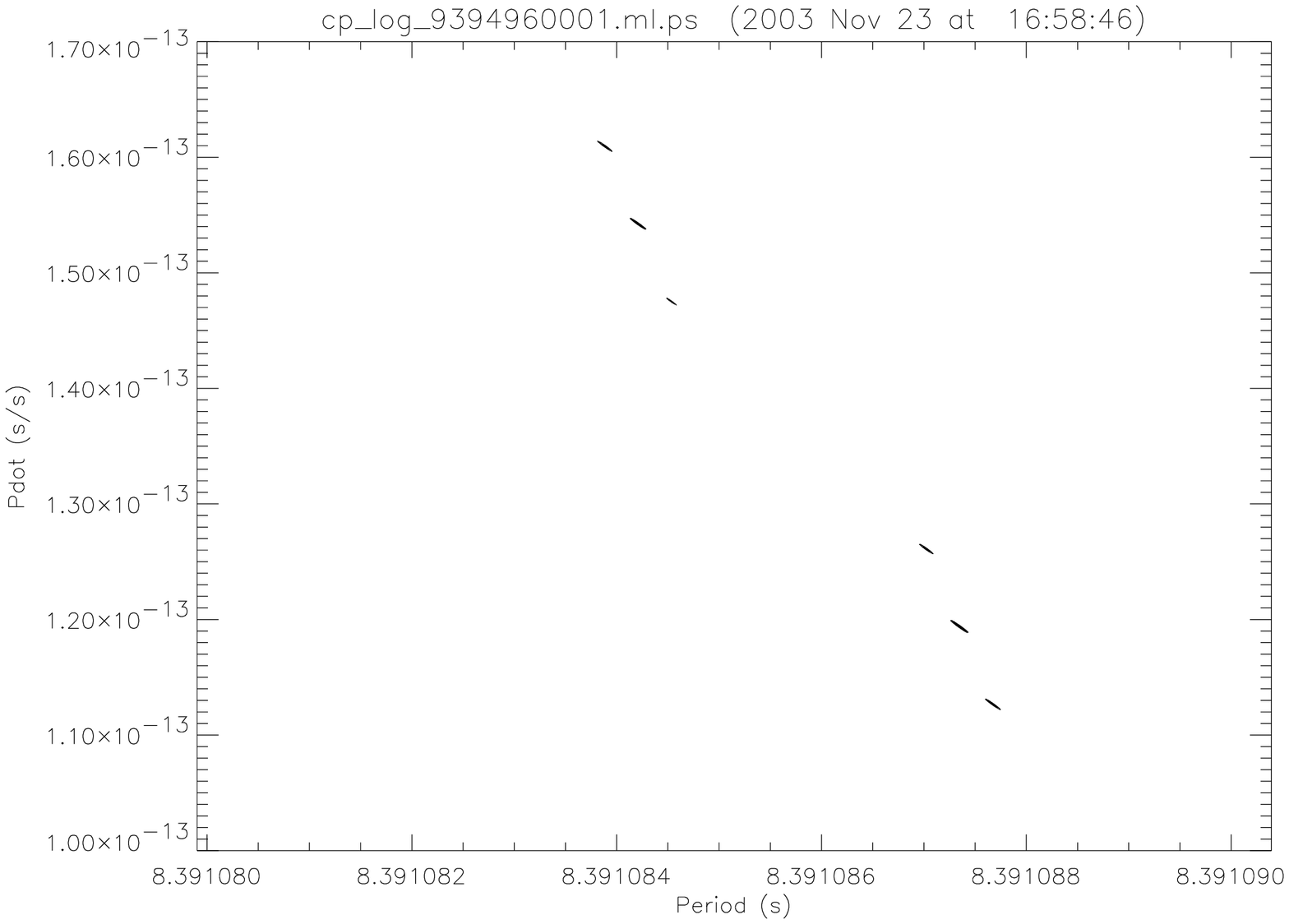}\\
\vspace*{10mm}
\includegraphics[trim=0.1cm 0 0 0,angle=90,width=\columnwidth]{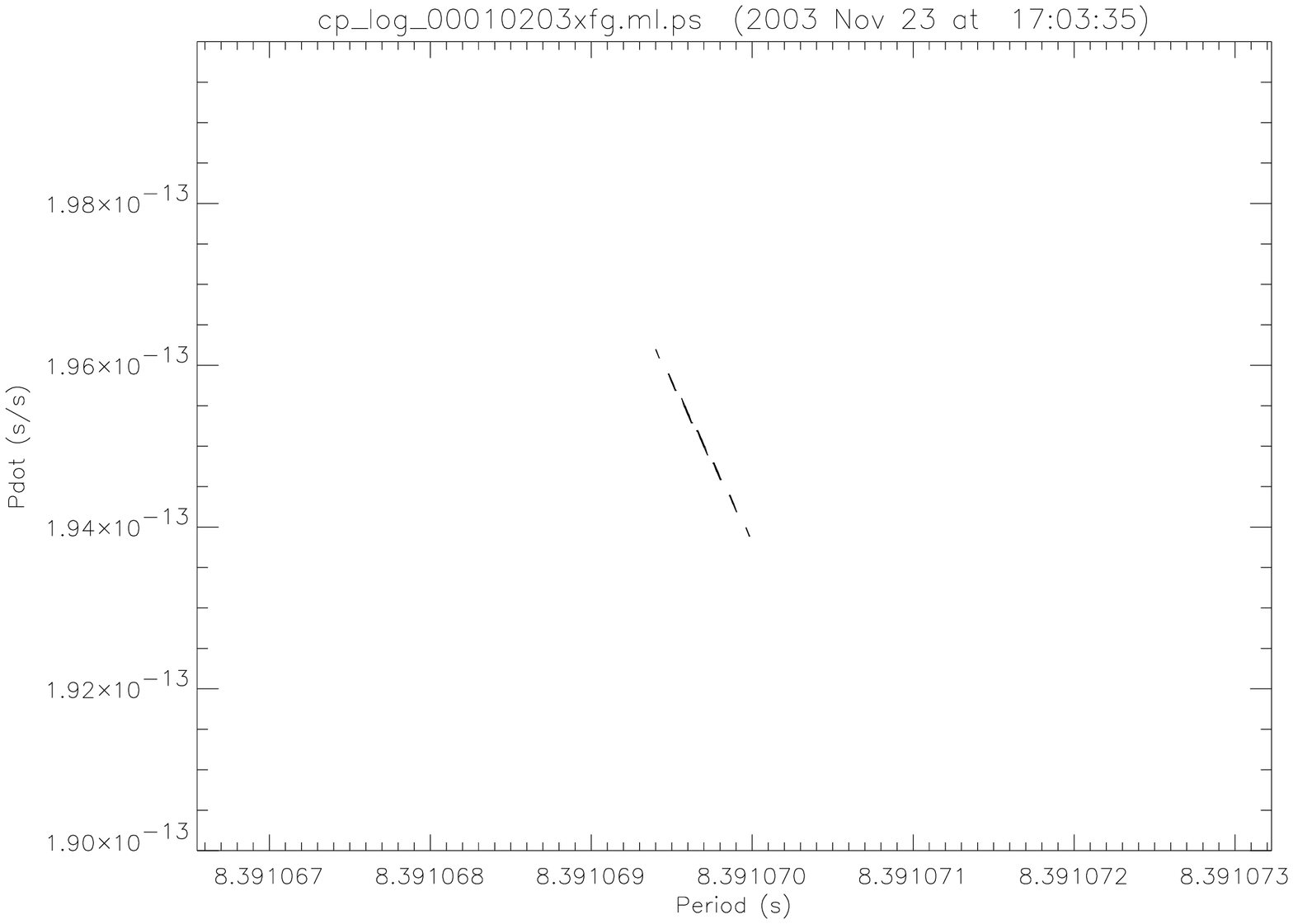}\\
\end{center}
\caption{The final 99\% confidence contours for the coherent analysis of
the combined R93, R94, R96, Ch00 and Ch01 datasets (top) and those for the
coherent analysis of the \xmm\/ datasets in more detail (bottom). In the 
bottom panel, the apparent aliases are the result of the contouring software.}
\label{fig:coherent_final}
\end{figure}

As discussed in \S~\ref{sec:indiv_sats}, we cannot 
find a consistent single $[P_0, \dot P]$ pair from
the analysis of the datasets from different satellites. However, there is
an overlap between the \ros\/ and \chandra\/ MLPs, therefore it is 
possible to compute a coherent MLP by restricting the analysis only to 
these data. The agreement in ToAs computed by K02
and ourselves here suggests that errors in the event timings are unlikely.  
Owing to the sparseness of the sampling, six aliases remain in the
coherent analysis. These all have $1.1\times10^{-13} < \dot P <
1.65\times10^{-13}$ and are shown in Figure~\ref{fig:coherent_final}(a).

The \xmm\/ dataset is the largest and provides highest S/N ratio MLPs. If
we take it at face value, a region with $\dot P \sim 1.95\times10^{-13}$ 
is identified in the MLP. This is shown in Figure~\ref{fig:coherent_final}(b).

\subsection{The incoherent analysis revisited}

In the light of the above uncertainties, it is necessary to examine
whether the incoherent analysis presented in Section~4 has been
compromised. We consider this unlikely, because the precision derived from
a coherent analysis of the time spanned by our data is nearly two orders
of magnitude greater than that which can be derived from the incoherent
analysis. The incoherent analysis is therefore unlikely to be affected by
the uncertainties at the accuracy required for it. The incoherent analysis
is also immune from many sources of error which might come to light only
when data from different epochs are combined, such as might arise for
light travel time corrections, or the transformation between time systems.

\section{Discussion}
\label{sect:disc}

\subsection{Interpretation of the spin down rate}
\label{sect:int}

We have presented a revision of the measure of the RX J0720.4--3125 spin
down rate published last year by Z02. By using additional \chandra \ and
\xmm \ datasets, we have been able to determine a {\it positive} spin down
rate of $\dot P = (1.4 \pm 0.6) \times 10^{-13}$~s/s at 99\%
confidence level, as derived from the incoherent analysis. More refined
values have been derived from a coherent analysis, separating
\ros\/ and \chandra\/ datasets from the \xmm \/ ones. However, a 
self-consistent phase-coherent solution based on the entire set of data 
from the three different satellites cannot be found. Due to these 
unresolved issues, in order to
discuss the physical implications of our measure we concentrate
on the result obtained from the incoherent analysis. Because of the 
larger numbers of datasets available since Z02 and K02, this measure of the
spin-down rate is now sufficiently constraining for our needs.

Despite the fact that the value of $\dot P$ reported here is higher than 
in
Z02, the two solutions are sufficiently similar that most of the
scientific conclusions published by Z02 remain unchanged.

First of all, our measure rules out the possibility that the source is
accreting from the interstellar medium, since this requires $\dot P <
10^{-15}$~s/s (see Z02, K02).

Furthermore, makes it unlikely that RX J0720.4--3125 is spinning 
down under propeller torques exerted by the interstellar medium. In Z02 
our argument
against this was based on the combination of two constraints between
magnetic field, density of the medium and star velocity. The first was
obtained from the expression of the propeller spin down, and the second by
imposing that the star has entered the propeller phase. The latter gives
$B_{12} < 25 {\sqrt n} v_{10}^{-3/2}$, where $n$ is the external density
in cm$^{-3}$,
$B_{12}= B
/(10^{12}~{\rm G})$ and $v_{10}$ is the star's velocity normalized to
10~km/s. For the former condition we used a propeller model as in Colpi et
al. (1998)
\begin{equation}
\dot P_{prop} \approx 10^{-8} n^{9/13} v_{10}^{-27/13} B_{12}^{8/13}
P^{21/13} \quad {\rm {s \over yr}}   \, ,
\label{prop}
\end{equation}
that, for $\dot P = (1.4 \pm
0.6) \times 10^{-13}$~s/s, gives
$B_{12} \approx (36-139) n^{-9/8} v_{10}^{27/8}$. Therefore, the two
conditions taken together
limit the star's velocity to
$v_{10} \la n^{1/3}$ ($n \sim 1$ for the interstellar medium). This
modest velocity was regarded as unlikely at the time of the Z02 paper, and 
it is now ruled out by the recent measure of a
proper motion of $97 \pm 12$~mas per year (Motch, Zavlin \& Haberl,~2003), 
which corresponds to a transverse velocity $V_T \sim 50 \times (
D/100 \/ pc)$~km/s. Since $v_{10}$ is likely to be greater than unity, the
propeller mechanism dominates only if the star is presently going through a
high density medium, with $n>100$. The possibility that RX J0720.4--3125
is passing through one of these high
density clouds (which are suggested to be present on small scales 
ranging from 10 to $10^6$~AU, see  Lauroesch \& Meyer~2000) has
been considered by Motch et al.~(2003).
However, their conclusion is that this scenario is problematic since it 
may imply changes in the X-ray flux and blackbody temperature (not 
observed so far) 
and it cannot account for a large $\dot P > 10^{-14}$~s/s.

We are then left with two possible explanations for the cause of
the observed spin down rate: propeller torque exerted by a
fossil disk or spin down due to emission of magnetic dipole radiation.
Although the first cannot be completely ruled out by our measure, 
and may better explain a high level of timing noise which is typical 
of accretors (Kaspi et al.~2001), we regard it as extremely unlikely. In 
fact, it requires two 
separate
mechanisms to explain the observed X-ray luminosity and spin down rate
(see Z02). Also, as discussed by Perna et al.~(2000) (see also Kaplan et
al.~2003), disk models fail in modeling the spectrum of the
optical counterpart.

\begin{figure}
\begin{center}
\includegraphics[width=\columnwidth]{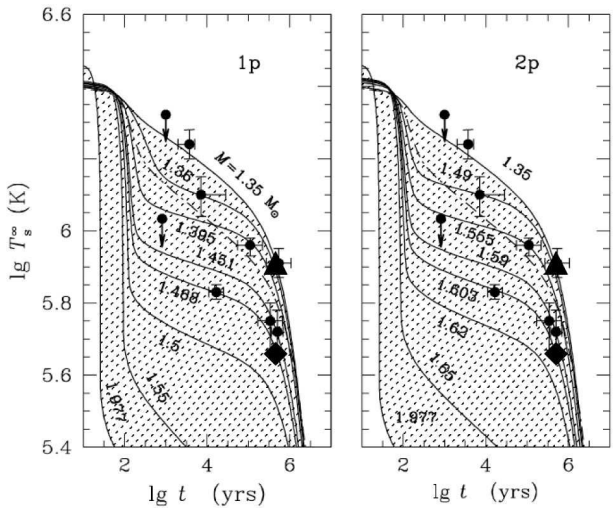}\\
\end{center}
\caption{The location of 9 isolated neutron stars in the temperature/age
diagram for two different models of proton superfluidity (1p and 2p) and
different masses (figure adapted from Yakovlev et al.~2003). 
In both panels, the diamond and the triangle mark the position of 
RX J0720.4--3125 as obtained assuming $T^\infty = 0.4
\times 10^6$~K and $T^\infty \sim 0.9
\times 10^6$~K, respectively (see text for details)}
\label{fig:cooling}
\end{figure}

By interpreting the spin down rate in terms of magnetic radiation losses
and assuming a dipolar field we can constrain the spin down age, $t_{sd}
= \dot P / \left ( 2P \right ) \sim (0.6-1.5) \times 10^6$~yr, and the
present value of the
magnetic field, $ B = (2.8-4.2) \times 10^{13}$~G.\footnote{As
in Z02, we reiterate that this must be regarded as an upper limit. A
twisted
magnetosphere may lead to a reduction up to an order of magnitude in the
inferred polar value of the magnetic field. The estimate of the
spin down age is unaffected (Thompson, Lyutikov \&
Kulkarni,~2001).}
The spin down age is consistent with the flight time of $10^6$~yrs from
the two birth places proposed by Motch et al.~(2003), the Sco OB2 complex
and the Vela OB2 + Trumpler 10 association.

\subsection{Location in the temperature/age diagram and estimate of the
mass}
\label{sect:cool}

The age derived from our
timing analysis can be then compared with the star's age as inferred by
the cooling curves, which is based on the spectral measure of the surface
temperature. 

Different theoretical cooling models have been presented in the 
literature. Roughly speaking, they predict
a two-fold behavior of the cooling curves: slow cooling for
low-mass neutron stars and fast cooling for stars with higher masses.
The transition between the two regimes can be more or less sharp, 
depending on the assumed superfluid properties of the neutron star 
interior. 
It has been realized (see Yakovlev et al.,~2003 for a recent 
discussion) that simple models which do not account for proton and
neutron superfluidity fail in explaining the surface temperatures observed
in many sources, unless there is a fine tuning in the masses of
objects such as Vela, Geminga and RX~J1856--3754, and all of them
have nearly the critical mass that bounds the (in this case, 
sharp) transition
between slow and fast cooling
regimes. Models with proton superfluidity
included do not require this
unlikely assumption and predict the existence of a region,
intermediate between the two regimes and relatively large in the 
parameter space, which is populated by medium
mass neutron stars (roughly between 1.4 and 1.65~\Msun).
Although the full picture only holds if,
at the same time, neutron superfluidity is rather weak, so far the 
latter models provide a better comparison with observations. It
is interesting that many neutron stars (as 1E~1207--52, Vela,
RX~J1856--3754, PSR~0656+14) have a surface temperature which falls in
such a transition region (see Figure~\ref{fig:cooling}, re-adapted from 
Yakovlev et al.,~2003).

If the origin of the X-ray flux from RX J0720.4--3125
is the thermal emission from the cooling neutron star, in first 
approximation the
surface temperature can be taken equal to the temperature of the 
blackbody fit, i.e. $T^\infty \sim 0.9
\times 10^6$~K (Haberl et al.,~1997; Paerels et
al.,~2001; Haberl et al.,~2004). This is likely to be an upper limit, 
since most detailed 
atmospheric models or fits which account for a broadband deviation from a 
blackbody tend to give a smaller temperature (see e.g. Pavlov et 
al.,~2002; Burwitz 
et al.,~2003). 
As we can see from Figure~\ref{fig:cooling}, this
gives
a maximum cooling age of $\sim 0.3 \times 10^6$~yr (for $M \sim 
1.35$~\Msun) which, given the
numerous uncertainties, may be considered as marginally compatible with
the spin-down age. Alternatively, Kaplan et al.~(2003) propose a
multicomponent fit to optical/X-ray data of RX J0720.4--3125 concluding 
that the X-ray emission at $\sim 0.9 \times 10^6$~K originates from hot
polar caps, while
the whole surface is cooler with a blackbody temperature of $T^\infty \sim
0.4 \times 10^6$~K. If this is the case, the spin down age is compatible
with RX J0720.4--3125 being a medium mass neutron star with $M \sim 
1.5-1.6$~\Msun, depending on the kind of proton superfluidity assumed
in the model (1p and 2p respectively).

It is worth noting that given the significant spread in the cooling curves in
Figure~\ref{fig:cooling} resulting from a strong mass dependence, an age as low
as $10^4$ yr (which may be possible if the magnetic field decays rapidly) is
also consistent with the observed surface temperature. We discuss these issues
in the next section.

\subsection{Implications for the past history of RX J0720.4--3125}
\label{sect:evol}

The considerations presented so far are based on the spin down age
$t_{sd}$, which is representative of the true
age of the source only in the case in which the magnetic field has remained
almost constant during the star evolution. The same condition applies for
the validity of the cooling curves mentioned above, which do not include
the extra input of energy released in the neutron star in case of field
decay. Taken face value, the timing parameters of RX J0720.4--3125 are
compatible either with those of a
radio-quiet
cooling pulsar with high (but not extreme) magnetic field or an old
magnetar, born with $B>10^{14}$~G and still kept warm by the decay of its
superstrong field. As in Z02, in order to discriminate between these
possibilities, we studied the evolutionary tracks on the $B, \dot P$
diagram. Three different mechanisms are
typically proposed for inducing field-decay: ambipolar diffusion in the
solenoidal or irrotational mode and Hall cascade (Goldreich \& 
Reisenegger~1992, Heyl \& Kulkarni~1998, Colpi, Geppert \& Page~2000). In 
reality all 
the three processes co-exist with different time scales, and each of them
may dominate the evolution depending on the istantaneous values of $B$,
$L$. In
absence of more detailed computations, Heyl \& Kulkarni~(1998) and Colpi,
Geppert \& Page~(2000) tentatively isolated the three processes and
proposed some simple, phenomenological rules to mimic the evolution in the
three regimes. As in Z02, we used as first approximations these
decay rules and re-computed the
source age, $\tau_d$, and the value of the magnetic field at the birth of
the neutron star, $B_0$ corresponding to our revised measure of $\dot P$.
By measuring $B_0$ in $10^{13}$~G, $P$ in
seconds, $\dot P$ in s/yr and the age in $10^6$~yr, the resulting
expressions are
\begin{equation}
B_0 = \left ( P \dot P \right ) ^{1/2} \left [ b^{\frac{\alpha-2}{2}} -
\frac{\alpha-2}{2} \frac{ a }{b} \left (P
\dot P \right) ^{\frac{\alpha-2}{2}} \left (P^2 - P_o^2 \right ) \right
]^{\frac{1}{2-\alpha}}
\end{equation}
\begin{equation}
\tau_d = \left ( a \alpha B_0^\alpha \right )^{-1}  \left \{ \left [ 1 -
\frac{2- \alpha}{2} \frac{a}{bB_0^{2-\alpha}}
\left (P^2 -
P_o^2 \right )
\right ]^{\frac{\alpha}{\alpha -2}} - 1 \right \}\, ,
\end{equation}
where $b\approx 3$, $P_0$ is the period at the star birth, and the
parameters $a, \alpha$ discriminate between the three decay laws (Colpi, 
Geppert \& Page,~2000). The
results are shown in Table \ref{decay}. In all cases the
star is assumed to be born with a period of 1 ms: results are not strongly
dependent on this exact value, provided it is less than the present
period\footnote{K02 discuss the possibility that RX J0720.4--3125 is an
example of the ``injection'' hypothesys (Vivekanand \& Narayan, 1981) and
is born with a long period, $P_0 \sim 8.3$~s. In this case, all three
decay laws predict an age of order $10^4$~yr.}.

\begin{table}
\begin{center}
\begin{tabular}{llll}
\hline
Solution & B-Decay Mechanism & $B_0$ & age \\
  &     &  $10^{13}$ G &  (years) \\
\hline
1& Hall Cascade & 120.2 & $3.5 \times 10^4$\\
1& Amb. diff., irrotational mode & 3.0 & $1.4 \times 10^6$\\
1& Amb. diff., solenoidal mode & 5.3 & $0.8 \times 10^6$\\
2& Hall Cascade & 121.6 & $2.3 \times 10^4$\\
2& Amb. diff., irrotational mode & 4.4 & $0.6 \times 10^6$\\
2& Amb. diff., solenoidal mode & 7.0 & $0.4 \times 10^6$\\
\hline
\end{tabular}
\caption{Predicted source age and primordial field for three different
mechanisms of decay, simulated as in Colpi et al.~(2000). Here $P =
8.391$~s, and we used different values of $\dot P$: $\dot P = 0.8 \times 
10^{-14}$~s/s
(solution `1') and $\dot P = 2 \times 10^{-13}$~s/s (solution `2').
These correspond
to the boundaries of the 99\% confidence range derived from our coherent
analysis. In all cases, the source is assumed to be born
with $P = 1
$~ms.}
\label{decay}
\end{center}
\end{table}
As we can see, the only scenario compatible with a
superstrong field at the star's birth ($B_0 \sim 10^{15}$~G) is that one
involving a very fast decay, i.e. the Hall cascade. In this
case the predicted star's age is $\sim 4 \times 10^4$~yr. Although such
a young age is only marginally compatible with the absence of a remnant,
it is not incompatible with that inferred from the cooling
curves (see Figure \ref{fig:cooling}) provided the mass is slightly larger. 
A slightly larger mass still is required if we allow for the extra-heating due
to $B$-decay from Hall cascade (Geppert \& Colpi, private communication).

On the other hand, both mechanisms involving
ambipolar diffusion predict a magnetic field which is relatively 
constant over the
source lifetime and close to the present value. The corresponding
star's age is between $\approx 0.4 \times 10^6$~yr and $\approx 1.4
\times 10^6$~yr, and is close to $\tau_{sd}$.

\section{Conclusions}
\label{sect:con}

We have eliminated the discrepancies between the calculated ToAs in the K02
paper and those derived from the same data analysed by Z02. This
invalidated our derived $[P_0, \dot P]$ solutions in Z02 (K02 have also
updated their ToAs). In order to correct these solutions, we have
reanalysed the datasets and taken the opportunity to incorporate
significant new datasets from both \chandra\/ and \xmm. At the same time,
we have revised the model used by Z02 to be fitted to the data to retain
the correct phase reference. 

Taken individually, the combination of old and new datasets allow a
significantly more accurate $[P_0, \dot P]$ solution to be derived than
that in K02 or Z02. This {\it establishes that RX J0720.4--3125 is spinning 
down}.

We have also attempted a phase-coherent analysis of the combined datasets,
as in Z02. However, we have been unable to find a consistent solution,
even when we analyse the data from different satellites separately. In
this case the \xmm\/ data are not consistent with the other data. We have
explored the possible reasons for this: they may be intrinsic to
RX J0720.4--3125 (in which case the constant $\dot P$ model is not
appropriate), or they may result from continuing timing errors in the data
we have not succeeded in identifying. We present the solutions from the
combined \ros\/ and \chandra\/ data and the \xmm\/ data separately.

The individual (incoherent) $[P_0, \dot P]$ solution is sufficiently
accurate to allows us to place strong constraints on the mechanism 
causing the spin-down. The large $\dot P \sim 10^{-13}$~s/s rules  
out accretion and propeller origins and points toward an interpretation in
terms of magneto-dipolar breaking. This in turn constrains the age of
RX J0720.4--3125 and its magnetic field strength.

For a magnetic field as strong as $B = (2.8-4.2) \times 10^{13}$~G, the
electron cyclotron line is unaccessible (it falls at about 350 keV).  However,
a relatively broad proton cyclotron absorption feature is predicted at $\approx
0.2-0.3$~keV (Zane et al.,~2001). The presence of a cyclotron absorption line
at the low energy of the sensitive energy band of the \xmm\/ EPIC instruments
has been indeed recently reported (see Haberl et al.,~2003a, 2004). The feature
may explain the observed deviations from a Planckian shape at these energies,
and possibly also the anticorrelation of the modulation of hardness ratio and
total X-ray intensity detected by {\xmm}\/ (Cropper et al.,~2001).

Our study allows a possible evolutionary link between dim INS and AXPs (which
relies on the interpretation of the former as warm-out magnetars) but only if
the INS have an age of $\sim10^4$ yr.  A more conservative interpretation is
that RX J0720.4--3125 was born with a strong, but not superstrong, field [$ B =
(2.8-4.2) \times 10^{13}$~G] which is compatible with those of the canonical
radio-pulsars. Similar conclusions have been argued also by K02, however their
measure of $\dot P$ has a large error and does not permit the discrimination
between this and the separate possibilities that RX J0720.4--3125 has a more
conventional field ($B\sim 10^{12}$~G) and lower spin-down rate ($\dot P \sim
10^{-15}$s/s) or is even spinning-up. Comparing with the radio-active pulsar
population, objects with $B>10^{13}$~G are rare, but their evidence is rapidly
growing (Camilo et al.,~2000). The parameters of RX J0720.4--3125 we have
derived  are not too dissimilar from those of PSR~J1814--1744, which has
$P\sim 4$~s and $\dot P \sim 7.4 \times 10^{-13}$~s/s.

One of the most striking mysteries about \ros \/ isolated neutron stars, AXPs,
soft-$\gamma$ repeaters and objects like Geminga is why they do not exhibit
radio emission.  As suggested by Motch et al.~(2003), who derived very similar
conclusions basing on the proper motion measurement, it could thus be that RX
J0720.4--3125 and several others apparently radio-quiet neutron stars are just
radio pulsars with off-axis beam that does not cross the Earth. The radio beam
narrows with increasing periods (Briggs 1990), making this explanation even
more plausible. On the other hand, evidence for a population of genuinely
radio-silent young neutron stars arises from population synthesis (Neuh\"auser
\& Tr\"umper,~1999; Gotthelf \& Vasisht, 2000; Popov et al.,~2000a,b and
references therein). The group of ROSAT INSs is very homogeneous and they are
all relatively close-by (within 300-400 pc), making unplausible that all their
beams are misaligned. At present, these two possibilities cannot be
distinguished. Detailed timing analysis of other \ros \/ INSs is crucial to
reach a more comprehensive understanding of the entire population.

\section{ACKNOWLEDGMENTS}

We are grateful to Uwe Lammers for his reprocessing of the \xmm\/ ODFs. We thank David Kaplan for several discussions regarding the ToAs and for pointing out the difference between the event timings in his and our BeppoSAX data. We thank Marco Feroci for processing the BeppoSAX data, and Tim Oosterbroek and GianLuca Israel for providing information on the intricacies of the time correction in the SAX DAS. We also thank Werner Becker for his insights into the possible sources of error in the \xmm\/ timing.

\end{document}